\documentclass[prx,twocolumn, aps,letterpaper, preprintnumbers,superscriptaddress ]{revtex4-2}
\usepackage{hyperref}
\usepackage{verbatim}
\usepackage{amsmath}
\usepackage{latexsym}
\usepackage{revsymb}
\usepackage{yfonts}
\usepackage{ifthen}

\usepackage{natbib}
\usepackage{amsfonts}
\usepackage{amsmath}
\usepackage{amssymb}
\usepackage{amsthm}
\usepackage{graphicx}
\usepackage{bm}
\usepackage{bbm}
\usepackage{epsfig,color,amssymb}
\usepackage{amsfonts}
\usepackage{amscd}
\usepackage{amsmath}
\usepackage{multirow}
\usepackage{chemarrow}
\usepackage{dcolumn}
\usepackage{bm}
\usepackage{graphicx}
\usepackage{enumerate}
\usepackage{epsfig}
\usepackage{subfigure}
\usepackage{xcolor}

\usepackage{ulem}
\usepackage{braket}
\usepackage{comment}
\usepackage{enumitem}
\usepackage{amsthm}
\usepackage{ragged2e}
\justifying\let\raggedright\justifying

\usepackage{booktabs}

\renewcommand{\emph}[1]{\textit{#1}}

\def\ket#1{\mathinner{|{#1}\rangle}}

\usepackage{amsfonts}
\usepackage{amsmath}
\usepackage{amsthm}
\usepackage{amscd}
\usepackage{amssymb}
\usepackage{amsxtra}
\usepackage{gensymb}
\usepackage{epstopdf}
\usepackage{color}
\usepackage{makecell}
\usepackage{multirow}
\usepackage{stackengine}
\usepackage{threeparttable}
\usepackage{upgreek}

\makeatletter
\renewcommand\@makefntext[1]{%
  \parindent 1em%
  \noindent\hbox to 1.8em{\hss$\@thefnmark$}#1
}
\makeatother

\makeatletter
\def\@fnsymbol#1{\ensuremath{\ifcase#1\or *\or \dagger\or \ddagger\or
   \mathsection\or \mathparagraph\or \|\or **\or \dagger\dagger
   \or \ddagger\ddagger \else\@ctrerr\fi}}
\makeatother

\makeatletter
\def\blfootnote{\xdef\@thefnmark{}\@footnotetext}
\makeatother

\hypersetup{colorlinks=true, linkcolor=blue, citecolor=red, urlcolor=blue}
\begin{document}
	\title{Experimental quantum secret sharing based on phase encoding of coherent states}

	\author{Ao Shen}\thanks{These authors contributed equally to this work}
 \affiliation{National Laboratory of Solid State Microstructures and School of Physics, Collaborative Innovation Center of Advanced Microstructures, Nanjing University, Nanjing 210093, China}
	\author{Xiao-Yu Cao}\thanks{These authors contributed equally to this work}
	\affiliation{National Laboratory of Solid State Microstructures and School of Physics, Collaborative Innovation Center of Advanced Microstructures, Nanjing University, Nanjing 210093, China}
        \author{Yang Wang}\thanks{These authors contributed equally to this work}
        \affiliation{National Laboratory of Solid State Microstructures and School of Physics, Collaborative Innovation Center of Advanced Microstructures, Nanjing University, Nanjing 210093, China}
        \affiliation{Henan Key Laboratory of Quantum Information and Cryptography, SSF IEU, Zhengzhou, China}
	\author{Yao Fu}\thanks{These authors contributed equally to this work}
	\affiliation{Beijing National Laboratory for Condensed Matter Physics and Institute of Physics, Chinese Academy of Sciences, Beijing 100190, China}
	\author{Jie Gu}\affiliation{National Laboratory of Solid State Microstructures and School of Physics, Collaborative Innovation Center of Advanced Microstructures, Nanjing University, Nanjing 210093, China}
        \author{\\Wen-Bo Liu}\affiliation{National Laboratory of Solid State Microstructures and School of Physics, Collaborative Innovation Center of Advanced Microstructures, Nanjing University, Nanjing 210093, China}
        \author{Chen-Xun Weng}\affiliation{National Laboratory of Solid State Microstructures and School of Physics, Collaborative Innovation Center of Advanced Microstructures, Nanjing University, Nanjing 210093, China}

	\author{Hua-Lei Yin}\email{hlyin@nju.edu.cn}
	\affiliation{National Laboratory of Solid State Microstructures and School of Physics, Collaborative Innovation Center of Advanced Microstructures, Nanjing University, Nanjing 210093, China}
	\author{Zeng-Bing Chen}\email{zbchen@nju.edu.cn}
	\affiliation{National Laboratory of Solid State Microstructures and School of Physics, Collaborative Innovation Center of Advanced Microstructures, Nanjing University, Nanjing 210093, China}

	\begin{abstract}	
		Quantum secret sharing (QSS) is one of the basic communication primitives in future quantum networks which addresses part of the basic cryptographic tasks of multiparty communication and computation. Nevertheless, it is a challenge to provide a practical QSS protocol with security against general attacks. A QSS protocol that balances security and practicality is still lacking. Here, we propose a QSS protocol with simple phase encoding of coherent states among three parties. Removing the requirement of impractical entangled resources and the need for phase randomization, our protocol can be implemented with accessible technology. We provide the finite-key analysis against coherent attacks and implement a proof-of-principle experiment to demonstrate our scheme's feasibility. Our scheme achieves a key rate of 85.3 bps under a 35 dB channel loss. Combined with security against general attacks and accessible technology, our protocol is a promising candidate for practical multiparty quantum communication networks.\\
		
	\end{abstract}

	\maketitle

\section{INTRODUCTION}
Quantum communication has attracted much attention~\cite{GisinThew-14,Kimble-11,MunroStephens-12,zhang2017quantum,liu2021practical,zhang2021external,liu2022fiber,sheng2022one,zhou2022one} due to its unconditional security during the communication process. Quantum communication has many branches, such as quantum key distribution~\cite{BENNETT20147,lucamarini2018overcoming,PRXliu21,zhou2021rate,ChenZhang-24,PhysRevLett.126.250502,liu2022decoy,PRXxie22,zhang2022revealing,PhysRevLett.128.180502,GU2022,PhysRevLett.128.110506}, digital signatures~\cite{PhysRevA.93.032316,RobertsLucamarini-16,yinnwac228}, quantum secret sharing (QSS)~\cite{hillery1999quantum,PhysRevLett.83.648} and so on. As an important branch of quantum communication, QSS is the quantum generalization of secret sharing. Secret sharing is an important primitive in quantum classical cryptography and was independently introduced by Shamir~\cite{shamir1979share} and Blakley~\cite{blakley1979safeguarding}. In a secret sharing scheme, a dealer splits a message into several parts and distributes each part to the corresponding player. In secret sharing, any unauthorized subset of players cannot reconstruct the message and the message can be reconstructed only when the authorized players cooperate. Since the security of classical secret sharing depends heavily on computational complexity, which has been proven vulnerable to future quantum computers, QSS is proposed to provide unconditional security based on the laws of quantum mechanics. The first QSS protocol was proposed by Hillery~\cite{hillery1999quantum} in 1999 by using the Greenberger-Horne-Zeilinger (GHZ) states. Since then, many QSS protocols have been proposed theoretically~\cite{bennett1992quantum,guo2003quantum,lance2004tripartite,xiao2004efficient,zhang2005multiparty,yan2005quantum,deng2006circular,qin2007cryptanalysis,markham2008graph,zhou2018quantum,gao2020deterministic,yang2021participant} and demonstrated experimentally~\cite{chen2005experimental,schmid2005experimental,gaertner2007experimental,bogdanski2008experimental,wei2013experimental,bell2014experimental,lu2016secret}. Although QSS has been studied extensively, it is still a challenge to provide a practical QSS protocol with security against general attacks. The first QSS protocol using GHZ states is not secure with a malicious player~\cite{qin2007cryptanalysis}. In addition, due to its insufficient transmission distance and a lack of efficient multiphoton sources, this QSS is highly impractical. To remove the requirement of the GHZ states, several single-qubit QSS schemes have been proposed~\cite{tavakoli2015secret,schmid2005experimental,hai2013experimental}. However, single-qubit schemes have drawbacks in their security~\cite{he2007comment,PhysRevLett.98.028902} and are vulnerable to Trojan horse attacks~\cite{xu2020secure}. In recent years, differential phase shift (DPS) QSS schemes~\cite{inoue2008differential,e23060716,gu2021differential} and round-robin (RR) QSS~\cite{wei2018quantum,gu2021secure} have been proposed. Using weak coherent states, DPS QSS schemes further simplify the experimental setup. Nevertheless, DPS QSS schemes~\cite{inoue2008differential,e23060716,gu2021differential} can defend against only individual attacks, such as photon number splitting and beam splitting attacks. RR QSS~\cite{wei2018quantum,gu2021secure} provides unconditional security against coherent attacks and removes the monitoring signal disturbance. However, implementation of the RR QSS lies in the realization of a variable-delay Mach-Zehnder interferometer, which improves the complexity of the experimental setup and constrains the practical application of this scheme. To date, a QSS protocol that balances security and practicality is still lacking.

Here, we present a QSS protocol that is secure against coherent attacks. By encoding logic bits with phase modulation on coherent states, our protocol removes the requirements for intensity modulation and phase randomization, which simplifies the experimental setup. Our protocol adopts the same remote single-photon interference method as twin-field quantum
key distribution to resist the Trojan horse attack in single-qubit QSS. Using the concentration inequality~\cite{curras2021tight,kato2020concentration} to consider statistical fluctuations, we provide the finite-key analysis against coherent attacks. In addition, we implement a proof-of-principle experiment of our protocol to demonstrate its feasibility in a plug-and-play system. We successfully generate secure key over various channel losses, up to 35 dB, under which a key rate of 85.3 bps can be obtained. Our protocol can be implemented by existing devices~\cite{PhysRevLett.123.100506,FangZeng-17,PittalugaMinder-18,cz-19,WangYin-20,zhou_22,clivati2022coherent} while achieving security against coherent attacks, which provides a possible solution for the application of QSS. Furthermore, as dishonest participants are allowed in QSS schemes, our protocol can be directly used for efficient quantum digital signatures~\cite{yinnwac228}. Addressing the bottleneck problems of QSS in both security and practicality, our QSS protocol paves the way to practical QSS networks.

\section{Protocol description}

\begin{figure}
	\centering
	\includegraphics[width=86mm]{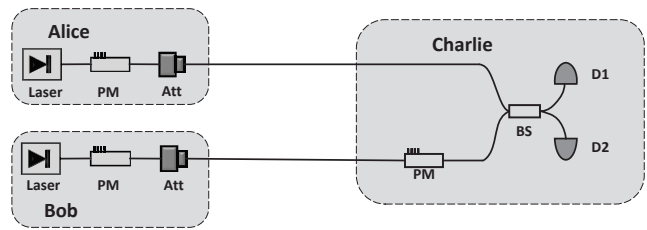}
	\caption{Setup of our quantum secret sharing protocol. Alice and Bob utilize lasers to emit optics pulses. They modulate the phase of their pulses with the phase modulator (PM) according to their basis choices and logic bits. After being dimmed by attenuators (Att), the pulses emitted by Alice and Bob are sent to Charlie. After adding a phase shift $\{0,\pi/2\}$ on pulses sent by Bob, Charlie lets the two pulses interfere with each other with a beam splitter (BS) and measures their phase difference with two single-photon detectors D1 and D2.
 } 
	\label{setup}
\end{figure}

The schematic diagram of our protocol is shown in Fig.~\ref{setup}. In our protocol, two symmetric distant players Alice and Bob send two weak coherent pulses, which are phase-encoded with logic bits and selected bases, to central dealer Charlie. After adding phase modulation to the optical pulse sent by Bob, Charlie performs an interference measurement on the two optical pulses sent by Alice and Bob and obtains the phase difference of the two pulses according to the clicking information of his detectors. Based on the measurement results and the basis choices of Alice, Bob and Charlie, each of the three parties in our protocol generates their own raw keys. Employing the same setup as the twin-field quantum key distribution, the key rate of our protocol scales with the square root of the total channel transmittance (between Alice and Bob) with single-photon interference. The detailed process of this QSS is as follows.

~\noindent{\it{1.~Preparation.}} In each turn, Alice (Bob) prepares a weak coherent pulse with intensity $\mu$ and selects $X$ and $Y$ bases with probabilities $p_x$ and $p_y=1-p_x$, respectively. In the $X$ basis, Alice (Bob) randomly modulates the phase of the pulse by $\{0,\pi\}$ and records her (his) logic bit as 0 (1) when the modulated phase is $0~(\pi)$. In the Y basis, Alice (Bob) randomly modulates the phase of the pulse by $\{\pi/2,3\pi/2\}$ and records her (his) logic bit as 1 (0) when the modulated phase is $\pi/2~(3\pi/2)$. Then, Alice and Bob send their pulses to the central dealer Charlie.

~\noindent{\it{2.~Measurement.}} Charlie imparts a phase 0 on the pulse sent by Bob with probability $p_x$ and imparts a phase $\pi/2$ with probability $p_y=1-p_x$. When the phase is $0~(\pi/2)$, Charlie records his basis as $X~(Y)$. Then Charlie performs an interference measurement on the two received pulses with a beam splitter. When the measurement is completed, Charlie records which detector clicks. If D1 (D2) clicks, then Charlie records
 his logic bit as 0 (1). If both detectors click, then Charlie randomly records his logic bit out of 0 or 1.

\begin{figure*}[t]
	\centering
	\includegraphics[width=\textwidth]{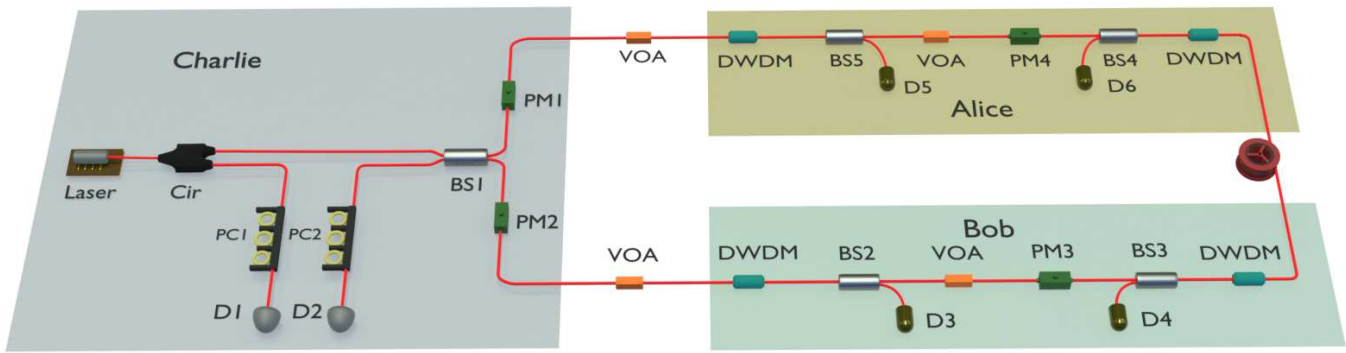}
	\caption{Experimental setup of our QSS protocol. Cir, circulator; PC, polarization controller; D1/D2, single-photon detector; BSi (i$\in\{1.2.3.4.5\}$), beam splitter ; PM1/PM2/PM3/PM4, phase modulator; VOA, variable optical attenuator; DWDM, dense wavelength division multiplexing. Note that the components including DWDMs and Di (i$\in\{3,4,5,6\}$) are not experimentally implemented due to resource limitations.
	}
	\label{exper}
\end{figure*}

~\noindent{\it{3.~Sifting.}} After sufficient turns of the above two steps, Alice, Bob and Charlie announce their basis choices. We denote the basis choice as $Z_i~(Z \in \{X,Y\}, i\in\{a,b,c\})$, where the subscript denotes who chooses the basis. If the basis choice of Alice, Bob and Charlie is one of $\{X_a,X_b,X_c\}$, $\{X_a,Y_b,Y_c\}$ and $\{Y_a,Y_b,X_c\}$, their logic bits are sifted to form their raw key bits. If the basis choice of Alice, Bob and Charlie is $\{Y_a,X_b,Y_c\}$, Alice and Bob sift their logic bits as their raw key bits, Charlie flips his corresponding logic bits to form his raw key bits. Otherwise, they discard their logic bits.

~\noindent{\it{4.~Parameter estimation.}} When the basis choice is one of $\{X_a,X_b,X_c\}$ and $\{X_a,Y_b,Y_c\}$, the raw key bits are used to form secure key bits and part of them are consumed to analyze the bit error rate $E_b^{\rm X}$. When the basis choice is one of $\{Y_a,X_b,Y_c\}$ and $\{Y_a,Y_b,X_c\}$, Alice, Bob and Charlie disclose their raw key bits to bound the phase error rate $E_p$.

~\noindent{\it{5.~Post-processing.}} Alice, Bob and Charlie conduct classical error correction and privacy amplification on the raw key bits, whose corresponding basis choice is $\{X_a,X_b,X_c\}$ or $\{X_a,Y_b,Y_c\}$, to distill the final keys.

To introduce our QSS protocol more explicitly, we demonstrate the bit correlation among Alice, Bob and Charlie. We denote $S_i \in \{0,1\}$ $ (i \in {a,b,c})$ as a classical bit, whose subscript denotes who holds this bit. In Step 1, the weak coherent states prepared by Alice (Bob) under the X basis can be expressed as $\ket{e^{iS_a\pi}\sqrt{\mu}}~\left(\ket{e^{iS_b\pi}\sqrt{\mu}}\right)$; the weak coherent states prepared by Alice (Bob) under the Y basis can be expressed as $\ket{e^{i(-S_a + \frac{3}{2})\pi}\sqrt{\mu}}~\left(\ket{e^{i(-S_b + \frac{3}{2})\pi}\sqrt{\mu}}\right)$. Since the period of the coherent state is $2\pi$, the weak coherent states prepared by Alice (Bob) under the Y basis can be equivalently expressed as $\ket{e^{i(S_a - \frac{1}{2})\pi}\sqrt{\mu}}~\left(\ket{e^{i(S_b - \frac{1}{2})\pi}\sqrt{\mu}}\right)$ after being imparted the phase $2(S_a - 1)\pi~\left(2(S_b-1)\pi\right)$. In Step 2, the phase modulation added by Charlie is 0 $(\pi/2)$ under the X (Y) basis. After Charlie modulates the phase of the pulse sent by Bob, we determine the phase difference between the two pulses sent by Alice and Bob. When the basis choice of Alice, Bob and Charlie is one of $\{X_a,X_b,X_c\}$, $\{X_a,Y_b,Y_c\}$ and $\{Y_a,Y_b,X_c\}$, the phase difference can be written as $\Delta\Phi = (S_b - S_a)\pi$, based on which we can calculate the bit correlation as $S_c= S_a \oplus S_b$. When the basis choice of Alice, Bob and Charlie is $\{Y_a,X_b,Y_c\}$, the phase difference can be written as $ \Delta \Phi = (S_b - S_a+1)\pi$. Note that Charlie flips his bits when the basis choice of Alice, Bob and Charlie is $\{Y_a,X_b,Y_c\}$; therefore, the bits of the three participants satisfy the same relationship that $S_c= S_a \oplus S_b$. In other words, Charlie's raw key bits are exclusive ORs of Alice's and Bob's raw key bits.

\section{Experimental demonstration}
Fig.~\ref{exper} shows the experimental setup
of our QSS protocol. This is a plug-and-play system consisting of a Sagnac interferometer. The frequency of the whole system is 100 MHz. The radio frequency signals provided for all PMs and synchronization signals of the whole system are from a high-speed arbitrary waveform generator with a sampling rate of 2.5 Gs/s (Tabor Electronics, P2588B).

Optical pulses are generated by a pulsed laser held by Charlie, with an extinction ratio greater than 30 dB. The temperature of the DFB laser
is modulated appropriately to provide optical pulses with a wavelength of 1550.12 nm. Charlie sends the optical pulses with a pulse width of 150 ps to Alice and Bob. The pulses go through a circulator (Cir) and are then separated by a 50:50 beam splitter (BS) into two identical pulses before entering the Sagnac loop. The Sagnac loop is used to stabilize the phase fluctuation of the channel automatically. The counterclockwise (clockwise) pulses are only modulated by Alice (Bob) without any other modulations. Since the pulses are generated by a third party, dense wavelength division multiplexings (DWDM), BSs and single-photon detectors should be added in the system, which are used for filtering and intensity monitoring, to prevent attacks from the injected pulses. Alice and Bob both select X and Y bases with probabilities $p_x=80\%$ and $p_y=20\%$. Specifically, for the $X$ basis, a 0 or $\pi$ phase will be randomly added, while for the $Y$ basis, a $\pi/2$ or $3\pi/2$ phase will be added.

Alice (Bob) sends pulses after phase modulation to Charlie. Alice's pulses interfere with Bob's pulses at Charlie's BS after passing through a variable optical attenuator (VOA) and a PM. The loss of communication channels is simulated with the VOA between Alice (Bob) and Charlie. Charlie adds only a phase 0 ($\pi/2$) on Bob's pulses with a probability $80\%$ ($20\%$) while adding an unmodulated phase 0 on Alice's pulses. Since the detection efficiencies of the two detectors are different, even if we randomly encode the pulse signals with logic bits 0/1, the number of 0/1 bits in the two detectors' detection results are different. To avoid the effect of such a difference on the final results, we added an extra 0 ($\pi$) phase to PM1, with a probability of 50\% (50\%), to balance the 0/1 bits of the two detectors' detection results. Two outputs of this BS are detected by two  superconducting nanowire single photon detectors, D1 and D2. The time windows of D1 and D2 are 1.7 ns and 2.1 ns, respectively, which are selected according to detection data. For D1, the detection efficiency $\eta_1$ is $86.5\%$ and the dark count rate $p_d^1=2.5\times{10}^{-8}$. For D2, $\eta_2=93.4\%$ and $p_d^2=3.5\times{10}^{-8}$. Detailed experimental data during implementation can be seen in Appendix.~\ref{exp_data}. Random numbers used in implementation are generated by Python's module random. 

\section{Security analysis} 
In this section, we present the security proof of our protocol. For simplicity, we assume that the source used in our protocol is perfect and only consider an internal eavesdropper when considering the security of the protocol. Since the information leaked to inside eavesdroppers is more than the information leaked to outside eavesdroppers, such an approximation does not compromise the security of our protocol. Then, we show that the security of our protocol against an inside eavesdropper is equivalent to the security of phase-encoded QKD. Based on the equivalence, the security of our QSS protocol can be demonstrated by utilizing the security analysis of phase-encoded QKD~\cite{lo2007security,koashi2009simple}. Furthermore, we present the finite-key analysis of our QSS protocol in Appendix.~\ref{fin_ana} and prove the security against coherent attacks in the finite-key regime.

\subsection{Security equivalence}
\begin{figure}
	\centering
	\includegraphics[width=86mm]{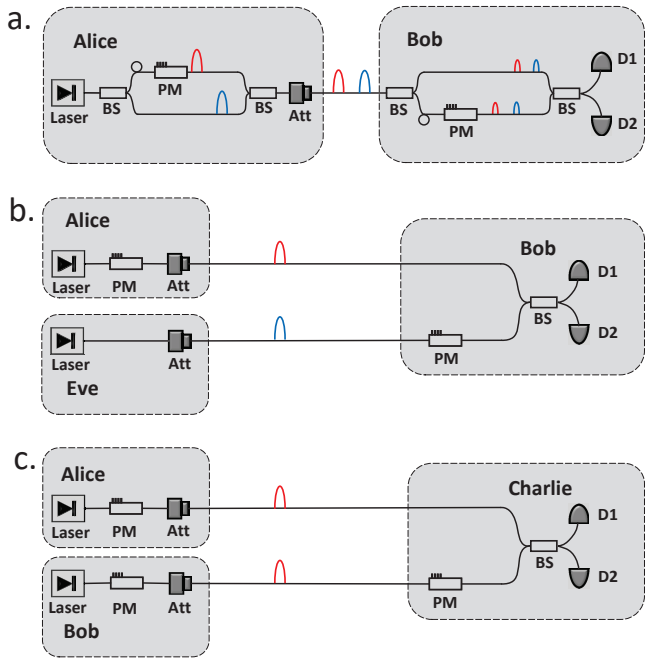}
	\caption{Configuration of phase-encoded QKD and our QSS protocol. (a) We present a typical schematic of phase-encoded QKD. In one period, Alice prepares a signal pulse (red) and a reference pulse (blue). After modulating the phase of the signal pulse, Alice sends the signal pulse and the reference pulse to Bob. Bob modulates the phase of the reference pulse and then makes the two pulses interfere to obtain their phase difference. (b) An equivalent scheme of phase-encoded QKD, in which eavesdropper Eve sends the reference pulses. (c) To build an equivalence between the security of phase-encoded QKD and our QSS protocol, we show the setup of our QSS protocol.} 
	\label{equivelence}
\end{figure}

In Fig.~\ref{equivelence}, we show the setup of phase-encoded QKD and our protocol. Fig.~\ref{equivelence}a shows the setup of phase-encoded QKD.
The laser source at Alice's site generates two pulses in one period, referring to one pulse as the signal pulse and the other pulse as the reference pulse. Alice modulates only the phase of the signal pulse and does not modulate the phase of the reference pulse. For the signal pulse, Alice selects one basis from X and Y. In the X basis, Alice phase modulates the signal pulse with $\{0,\pi\}$. In the Y basis, Alice phase modulates the signal pulse with $\{\pi/2,3\pi/2\}$. When the pulses reach Bob's site, Bob selects one basis from X and Y for each pulse. Under the X(Y) basis, Bob imparts a phase 0 ($ \pi/2 $) to the reference pulse. Then, Bob makes the two pulses interfere and measures them with two detectors D1 and D2.

As an intermediate step, we consider an equivalent scheme (Fig.~\ref{equivelence}b) of phase-encoded QKD (Fig.~\ref{equivelence}a). In this scheme, Alice sends signal pulses in the same way as phase-encoded QKD, while the eavesdropper Eve sends the reference pulses. Since the security of phase-encoded QKD can be proven even if the eavesdropper knows the phase of the reference pulse~\cite{lo2007security}, the two schemes are equivalent from a security perspective. To build the equivalence of security between phase-encoded QKD and our QSS protocol, we present the setup of our protocol in Fig.~\ref{equivelence}c. We consider the case with an inside eavesdropper and assume that Bob is the inside eavesdropper. Comparing Fig.~\ref{equivelence}b and Fig.~\ref{equivelence}c, we find that the difference between the two schemes is that Eve, the eavesdropper in phase-encoded QKD, sends reference pulses, while Bob, the eavesdropper in our QSS scheme, sends phase-encoded pulses. Since the phase information imparted on the pulse sent by the eavesdropper has no influence on the security of phase-encoded QKD, the security of phase-encoded QKD can be proven to be equivalent to the security against an inside eavesdropper of our QSS protocol.

\subsection{Inside eavesdropper}
Without loss of generality, let us assume that Bob is an internal eavesdropper. We introduce the concept of ``quantum coin'' to bound the information leakage of the signal sent by Alice. The imbalance $\Delta$ of the ``quantum coin'' is related to the basis dependence of the signal of Alice. Alice's basis-dependent entangled states can be expressed as
\begin{equation}
	\begin{aligned}
	    &\ket{\Psi_x}= (\ket{0_X} \otimes \ket{\alpha} + \ket{1_X} \otimes \ket{-\alpha})/\sqrt{2},\\
		&\ket{\Psi_y}= (\ket{1_Y} \otimes \ket{i\alpha} + \ket{0_Y} \otimes \ket{-i\alpha})/\sqrt{2},\\
	\end{aligned}
\end{equation}
where ${\ket{0_X},\ket{1_X}}$ are eigenstates of the Pauli operator $\sigma_x$ and ${\ket{0_Y},\ket{1_Y}}$ are eigenstates of the Pauli operator $\sigma_y$. To quantify the basis dependence of Alice's signal pulses, we can relate the basis dependence with the ``balance'' of a ``quantum coin'' in an equivalent protocol~\cite{lo2007security}. In this equivalent protocol, Alice measures the quantum coin in the basis ${\ket{0_Z},\ket{1_Z}}$ to determine whether Alice sends state $\ket{\Psi_x}$ or state $\ket{\Psi_y}$. The joint state of the quantum coin state and Alice source state can be taken to be
\begin{equation}
	\begin{aligned}
		\ket{\Phi} = \sqrt{p_x} \ket{0_Z} \otimes \ket{\Psi_x} + \sqrt{p_y} \ket{1_Z} \otimes \ket{\Psi_y},\\
	\end{aligned}
\end{equation}
where $p_x$ and $p_y$ are the probability with which Alice encodes her signal in the X basis and in the Y basis, respectively. Furthermore, we assume that the measurement of the coin is delayed until after the eavesdropper is finished eavesdropping on the signals. In our QSS protocol, when Alice, Bob and Charlie's basis choice is $\{Y_{a},X_{b},Y_{c}\}$ or $\{Y_{a},Y_{b},X_{c}\}$, their logical bits are used to calculate phase errors. The phase error rate $E_p$ can be bounded by a function of $E_b^{\rm Y}$ and $\Delta$ and expressed as
\begin{equation}
	\begin{split}
		E_p = ~& E_b^{\rm Y} + 4\Delta(1-\Delta)(1-2E_b^{\rm Y})  \\
		& + 4 (1-2\Delta)\sqrt{\Delta(1-\Delta)E_b^{\rm Y}(1-E_b^{\rm Y})},\\
	\end{split}
 \label{eq3}
\end{equation}
where $E_b^{\rm Y}$ is the bit error rate, which is calculated by using the key bits corresponding to basis choice $\{Y_{a},X_{b},Y_{c}\}$ or $\{Y_{a},Y_{b},X_{c}\}$, and $\Delta$ quantifies the basis dependence of Alice's signals. $\Delta$ can be calculated by              
\begin{equation}
	\begin{aligned}
		1-2Q_\mu\Delta = \langle\Psi_y|\Psi_x\rangle,\\
	\end{aligned}
\end{equation}
where $Q_\mu$ is the total gain of the protocol.

In the asymptotic case, the final key rate of our QSS protocol can be expressed as
\begin{equation}
	\begin{aligned}
		R = Q_\mu[1-f_e H(E_{\rm b}^{\rm X}) - H(E_{\rm p})],
	\end{aligned}
\end{equation}
where $f_e$ is the error-correction efficiency; $E_{\rm b}^{\rm X}$ is the bite error rate of the raw key bits whose corresponding basis choice is $\{X_{a},Y_{b},Y_{c}\}$ or $\{X_{a},X_{b},X_{c}\}$; $H(x) = -x\log_{2}{x} - (1-x)\log_{2}{(1-x)}$ is Shannon entropy. For Charlie's two detectors, we assume that the dark count rate of both is $p_{\rm d}$. $Q_\mu$ is the gain corresponding to the basis choice $\{X_{a},Y_{b},Y_{c}\}$ and $\{X_{a},X_{b},X_{c}\}$. With a light intensity $\mu$ of weak coherent states, $Q_{\mu}$ is given by $Q_\mu = (1-p_{\rm d})[1 - (1-2p_{\rm d}) e^{-2\mu \eta}]$. The bit error rate $E_{\rm b}^{\rm X}$ is given by $E_{\rm b}^{\rm X} Q_\mu = e_{\rm d} (1 -p_{\rm d}) [1 - (1- p_{\rm d}) e^{-2\mu \eta}] + (1 - e_{\rm d}) p_{\rm d} (1 - p_{\rm d}) e^{-2\mu \eta}$, where $e_{\rm d}$ is the misalignment error rate of the detectors.
	
\section{Results}\label{final}
Here, to evaluate the performance of our protocol, we conduct the finite key rate of our QSS protocol. In the above section, we obtain the information leakage to malicious Bob. Since the information leakage to malicious Bob is greater than that to external Eve, we can consider the information leakage in our protocol to be the information leakage to malicious Bob. Alice and Bob are two symmetric participants in our protocol, so we take the total distance between Alice and Bob as $L$ in the simulation. Based on such a configuration, the channel transmittance $\eta$ becomes $\eta_{\rm d} \times 10^{-\alpha L/20}$, where $\eta_{\rm d}$ is the detection efficiency of Charlie's detectors and $\alpha$ is the attenuation coefficient of the ultra-low fiber.

\subsection{Simulation results}

The finite key rate of our QSS protocol can be given by
\begin{equation}
	\begin{aligned}
		l  =&~ n_{x}\bigg[1 - H(\overline{E}_{\rm p}) - {\rm leak}_{\rm EC}  \\
		& - \frac{1}{n_{x}}\log_{2}{\frac{2} {\epsilon_{\rm c}}} - \frac{1}{n_{x}}\log_{2}{\frac{1}{4\epsilon^2_{\rm PA}}}\bigg],
	\end{aligned}
\end{equation}
$\epsilon_{\rm c}$ and $\epsilon_{\rm PA}$ represent the probability of failure in error correction and privacy amplification, respectively. $n_{x}$ denotes the number of raw key bits used to generate the secret key bits whose corresponding basis choice is $\{X_{a},X_{b},X_{c}\}$ or $\{X_{a},Y_{b},Y_{c}\}$. $ {\rm leak_{\rm EC}} =  f_e H(E_{\rm b}^{\rm X})$ represents the fraction of bits
\begin{figure}
	\centering
	\includegraphics[width=86mm]{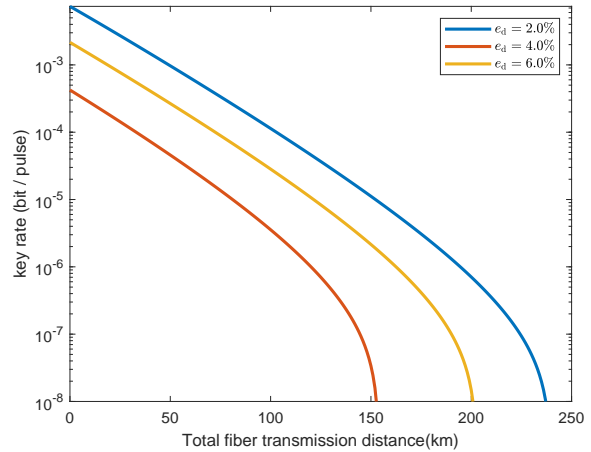}
	\caption{Simulation key rate with variant misalignment rates. Simulation results for finite final key rate as a function of fiber transmission length with a fixed key length $N=10^{10}$ and misalignment rate $e_{\rm d} = 2\%,4\%,6\%$.}
	\label{key}
\end{figure}
 consumed for error correction, where $f_e$ is the error-correction efficiency and $E_{\rm b}^{\rm X}$ represents the bit error rate of the raw key bits whose corresponding  basis choice is $\{X_{a},X_{b},X_{c}\}$ or $\{X_{a},Y_{b},Y_{c}\}$. $\overline{E}_{\rm p}$ is the upper bound of the phase error rate, whose detailed formulas are given in Appendix.~\ref{fin_ana}.

Assuming the size of the finite key is $N=10^{10}$, we utilize the genetic algorithm to optimize the finite key rate under a certain distance with the misalignment rate $e_{\rm d} = 2\%,4\%,6\%$. Other simulation parameters are summarized as follows: $\eta_{\rm d} = 56\%$, $p_{\rm d} = 10^{-8}$, $\alpha = 0.167$, and $f_e = 1.16$. We show our simulation result in Fig.~\ref{key}. As shown in Fig.~\ref{key}, our QSS protocol achieves a transmission distance of more than 230 km when the misalignment rate $e_{\rm d} = 2\%$. Furthermore, our protocols show tolerance for high misalignment error rates. With an error rate of 6 \%, a transmission distance of 150 km can still be obtained.

\subsection{Experimental results}

\begin{table*}[htbp]
	\footnotesize
	\caption{Experimental data. We demonstrate the feasibility of our protocol under different channel losses. N is the number of pulses sent, $\mu$ is the higher intensity of the pulses sent by Alice and Bob, $E_{\rm b}^{\rm X}$ and $E_{\rm b}^{\rm Y}$ are the experimental quantum bit error rates in the $X$ basis and $Y$ basis, $n_x$ and $n_y$ are the number of clicks in the $X$ basis and $Y$ basis, $R$ is the key rate.}
	\label{result_table}
	\tabcolsep 18pt 
	\begin{tabular*}{\textwidth}{cccccccc}
		\toprule
		
		Loss & N & $\mu$ & $E_{\rm b}^{\rm X}$ & $E_{\rm b}^{\rm Y}$ & $n_x$ & $n_y$ &  $R$ \\ \hline
		
		20 dB & $10^{10}$ &  5.8$\times 10^{-3}$ & 0.16\% & 0.09\% & 2776599 & 315364 & 7.51$\times 10^{-5}$ \\
		30 dB&$10^{10}$&  1.6$\times 10^{-3}$ &0.19\%& 0.09\%  & 239619 & 27474  &4.67$\times 10^{-6}$\\
		35 dB&$10^{10}$&  8.6$\times 10^{-4}$&0.30\% &0.30\%  & 73954 & 8346 & 8.53$\times 10^{-7}$ \\ 
		
		\bottomrule
	\end{tabular*}
\end{table*}

We implement our protocol in the finite-key regime over various losses of 20, 30, and 35 dB. Under various losses, we performed this protocol with optimized intensity and a fixed ratio of X basis to Y basis (4:1).

\begin{figure}
	\centering
	\includegraphics[width=86mm]{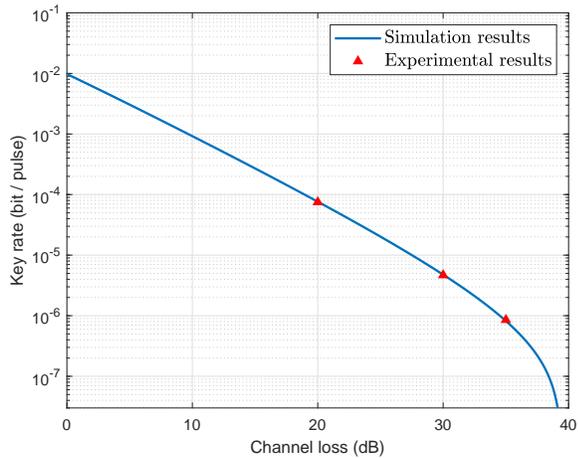}
	\caption{Secure key rate of our QSS scheme. The secure key rates are plotted against the total loss from Alice to Bob with the total number of pulses sent by Alice or Bob $N = 10^{10}$. The triangular-type dots correspond to the experimental results with fiber transmitting losses of 20 dB, 30 dB, and 35 dB.}
	\label{rate_exp}
\end{figure}

The experimental results we obtained are listed in
Table~\ref{result_table} and shown in Fig.~\ref{rate_exp}.  Given the 100-MHz repetition rate, our protocol can
achieve a secure key rate of 85.3 bps with a channel loss of over 35 dB, allowing it to be deployed over 175 km with available technologies. A secure key rate of 7.51 kbps was generated at 20 dB ($\sim$ 100 km) while at 30 dB ($\sim$ 150 km), it was 467 bps. Note that the intensities of pulses sent by Alice and Bob are different
 due to the different insertion losses of Charlie's PMs and the difference has little influence on results. We used the larger value of the pulse intensity to calculate the final result, which results in a lower key rate.

As a tradeoff for automatic phase stabilization, long-distance optical fibers were not used in the system and the two users were connected with a one-meter-long fiber. Two users in the plug-and-play system are unable to resist Trojan horse attacks. Additionally, due to the lack of DWDMs and photodiodes, we did not actually monitor the intensity and filter pulses, which should be addressed to prevent potential attacks. However, instead of establishing a complete system with all necessary elements, we concentrate on demonstrating the feasibility of our protocol. The lacking devices can be directly added to our system without invalidating the obtained experimental results. Employing the developed technology in twin-field QKD, such as phase-locking and phase-tracking ~\cite{PhysRevLett.123.100506,FangZeng-17,PittalugaMinder-18,cz-19,WangYin-20,zhou_22,clivati2022coherent}, a scheme with two independent users and long-distance fibers can be realized.

\section{Conclusion}\label{conclu}
In summary, we propose a quantum secret sharing protocol that balances security and practicality. The use of weak coherent states in our protocol removes the requirements for impractical entangled sources. Our protocol removes the requirements for intensity modulation and phase randomization, which avoids pattern effects in experimental implementation and enables commercial applications of our protocol. In addition, our protocol uses the same device as twin-field QKD~\cite{PhysRevLett.123.100506,FangZeng-17,PittalugaMinder-18,cz-19,WangYin-20,zhou_22,clivati2022coherent} and can therefore be implemented with existing devices. Furthermore, our QSS protocol can be directly used for efficient quantum digital signatures~\cite{yinnwac228}. Using the concentration inequality~\cite{curras2021tight,kato2020concentration} to consider statistical fluctuations, we provide a finite-key analysis against coherent attacks for our QSS protocol. In the finite-key regime, our QSS protocol achieves a theoretical transmission distance over 230 km under a small misalignment error rate while showing tolerance for high misalignment rates, which makes field tests possible.

Furthermore, we experimentally demonstrate our protocol in a plug-and-play system. Our protocol can be demonstrated over 35 dB with the ability to resist coherent attacks. Without intensity modulation and phase randomization, our protocol can still achieve a key rate of 85.3 bps under a 35 dB channel loss, which outperforms other experimental implementations of QSS protocols. In addition, we also achieve key rates of 7.51 kbps and 0.467 kbps at 20 dB and 30 dB, respectively. Combined with high security and simple apparatus requirements, our protocol paves the way for secure multiparty communication in future quantum networks.

\appendix
\section*{Acknowledgments}
This study was supported by the National Natural Science Foundation of China (No. 12274223), the Natural Science Foundation of Jiangsu Province (No. BK20211145), the Fundamental Research Funds for the Central Universities (No. 020414380182), the Key Research and Development Program of Nanjing Jiangbei New Aera (No. ZDYD20210101),  the Program for Innovative Talents and Entrepreneurs in Jiangsu (No. JSSCRC2021484), and the Program of Song Shan Laboratory (Included in the management of Major  Science and Technology Program of Henan Province) (No. 221100210800-02).

\section{Detailed experimental data}\label{exp_data}

Detailed experimental data are presented in Table~\ref{EXP_DET}, including the number of all detection events n and the number of detection events under different added phases. The number of detection events under different phases added is named as ``Detected ABC", where ``A'' (``B'' or ``C'') means an A (B or C) phase was added on the pulses by Alice (Bob or Charlie) . The pre-calibrated losses are depicted in Table~\ref{EFF_OPT}. The elements include PMs, PCs, Cir, and BSs. The results are given for each output (D1/D2) as appropriate.

\begin{table}[htbp]
\centering
	\footnotesize
	\caption{Efficiencies of the elements of the measurement station.}\label{EFF_OPT}
	\doublerulesep 0.1pt \tabcolsep 13pt 
	\begin{tabular}{cc}
		\toprule
		Optical devices &  Insertion loss \\ 
		\midrule
		Cir 2$\rightarrow$3  & 0.62 dB \\ 
		BS-A  & 0.69 dB \\ 
		BS-B  & 0.70 dB \\ 
		PM-A  & 1.93 dB \\ 
		PM-B  & 2.01 dB \\ 
		$\rm{PC_1}$   & 0.18 dB \\
		$\rm{PC_2}$   & 0.16 dB \\
		\bottomrule
	\end{tabular}
\end{table}

\begin{table*}[t]
	\centering
	\footnotesize
	\caption{Detailed experimental data under different channel losses.}
	\label{EXP_DET}
	\tabcolsep 22pt 
	\begin{tabular*}{\textwidth}{ccccccc}
		\toprule
		Channel loss & \multicolumn{2}{c}{20 dB} & \multicolumn{2}{c}{30 dB} & \multicolumn{2}{c}{35 dB} \\ 
		\midrule
		
		n & \multicolumn{2}{c}{5135663} & \multicolumn{2}{c}{442749} & \multicolumn{2}{c}{136412} \\ \hline
		Detector & D1 & D2& D1 & D2& D1 & D2\\
		Detected 000  &567134 &1426 &48972  &127 &15129 &74 \\ 
		Detected 0$\pi$0 &1085 & 762615 &104 & 66325 &48&20236 \\ 
		Detected $\pi$00 &725&718749 &92&61975 &39 &19103\\ 
		Detected $\pi$$\pi$0 &581843 & 1002 &49848&111 &15535 &51\\
		Detected 0$\frac{\pi}{2}$$\frac{\pi}{2}$ &29 &36425 &4&3085 &2 &964\\ 
		Detected 0$\frac{3\pi}{2}$$\frac{\pi}{2}$ &33149&87 &2785&9 &886 &4\\ 
		Detected $\pi$$\frac{\pi}{2}$$\frac{\pi}{2}$ &31191&41 &2648 &5 & 817&4 \\ 
		Detected $\pi$$\frac{3\pi}{2}$$\frac{\pi}{2}$ &21&41077 &5 &3524 & 2&1060\\ 
		Detected $\frac{\pi}{2}$$\frac{\pi}{2}$0 & 37300&30 & 3261& 0 & 1032&1\\ 
		Detected $\frac{\pi}{2}$$\frac{3\pi}{2}$0 &22&40704 &0 &3515 &1&1065 \\
		Detected $\frac{3\pi}{2}$$\frac{\pi}{2}$0 &42&44354 &4 &3864 &4&1222 \\
		Detected $\frac{3\pi}{2}$$\frac{3\pi}{2}$0 &39512&50 & 3462& 6 & 1055&6\\ 
		Detected $\frac{\pi}{2}$0$\frac{\pi}{2}$ &37565&45 &3231 &5 &975&2 \\ 
		Detected $\frac{\pi}{2}$$\pi$$\frac{\pi}{2}$& 38&37342&3&3199 & 6&959\\
		Detected $\frac{3\pi}{2}$0$\frac{\pi}{2}$ & 28 & 38821&5&3412 &3 &987\\ 
		Detected $\frac{3\pi}{2}$$\pi$$\frac{\pi}{2}$ &39469 &42& 3505 & 2&1026&2 \\ 
		
		\bottomrule
	\end{tabular*}
\end{table*}


\section{Finite-key analysis}\label{fin_ana}

Here, we provide the finite-key analysis of our protocol. Using the concentration inequality~\cite{kato2020concentration} in Ref.~\cite{curras2021tight}, we provide security against coherent attacks. The formula of the finite key rate can be expressed as

\begin{equation}
	\begin{aligned}
		l  = &~ n_{x}\bigg[1 - H(\overline{E}_{\rm p}) - {\rm leak}_{\rm EC}   \\
		& - \frac{1}{n_{x}}\log_{2}{\frac{2} {\epsilon_{\rm c}}} - \frac{1}{n_{x}}\log_{2}{\frac{1}{4\epsilon^2_{\rm PA}}}\bigg],
	\end{aligned}
\end{equation}
which is proven to be $\epsilon_{\rm c}$-correct and $\epsilon_{s}$-secure, with $\epsilon_s = \sqrt{\epsilon} + \epsilon_{\rm PA}$. $\epsilon$ is the failure probability associated with the estimation of the phase error rate. $\epsilon_{\rm c}$ and $\epsilon_{\rm PA}$ represent the probability of failure in error correction and privacy amplification, respectively. $n_{x}$ denotes the number of raw key bits used to generate the secret key bits whose corresponding basis choice is $\{X_{a},X_{b},X_{c}\}$ or $\{X_{a},Y_{b},Y_{c}\}$. $ {\rm leak_{\rm EC}} =  f_e H(E_{\rm b}^{\rm X})$ represents the fraction of bits consumed for error correction, where $f_e$ is the error-correction efficiency and $E_b^X$ represents the bit error rate of the raw key bits whose corresponding basis choice is $\{X_{a},X_{b},X_{c}\}$ or $\{X_{a},Y_{b},Y_{c}\}$. In addition, we apply statistical fluctuations to calculate the upper bound of the observed phase error rate $\overline{E}_{\rm p}$. When considering statistical fluctuations, we use the concentration inequality~\cite{kato2020concentration,curras2021tight} to estimate the upper bound of the deviation between a sum of correlated random variables and its expected value. The concentration inequality is tighter than the widely employed Azuma's inequality~\cite{kazuoki1997Weighted} and proves the security against coherent attacks.

We define $\xi_{1},...,\xi_{n}$ to be a sequence of Bernoulli random variables and define $\Lambda_j$ to be the sum of these random variables, i.e. $\Lambda_j=\sum_{u=1}^{j}\xi_{u}$. Let $\mathcal{F}_j$ denote the $\sigma$-algebra generated by $\{\xi_{1},...,\xi_{n}\}$ that is the natural filtration of those Bernoulli random variables. Let $\epsilon_a$ denote the failure probabilities for the concentration bound for sums of dependent random variables. Using the results in Refs.~\cite{curras2021tight,kato2020concentration}, we find that for any $b>0$
\begin{equation}
	\begin{aligned}
		{\rm Pr}\left [\Lambda_n- \sum \limits_{u=1}^{n}{\rm Pr}(\xi_{u}=1|\mathcal{F}_{u-1})\ge b\sqrt{n}\right ] \leq {\rm exp}[-2b^2],\\
		{\rm Pr}\left [\sum \limits_{u=1}^{n}{\rm Pr}(\xi_{u}=1|\mathcal{F}_{u-1})-\Lambda_n \ge b\sqrt{n}\right ] \leq {\rm exp}[-2b^2].\\
	\end{aligned}
	\label{concen1}
\end{equation}
Equating the right-hand sides of eq.~(\ref{concen1}) to $\epsilon_{a}$ and solving for $b$, we find a simple bound of concentration inequality, which can be expressed as
\begin{equation}
	\begin{aligned}
		\sum \limits_{u=1}^{n}{\rm Pr}(\xi_{u}=1|\xi_{1},...,\xi_{u-1}) \leq \Lambda_n + \Delta_{c},\\
		\Lambda_n \leq \sum \limits_{u=1}^{n}{\rm Pr}(\xi_{u}=1|\xi_{1},...,\xi_{u-1}) + \Delta_{c},
	\end{aligned}
	\label{concen2}
\end{equation}
where $\Delta_c = \sqrt{\frac{1}{2}n\ln\epsilon_{\rm a}^{-1}}$ and $\epsilon_a$ is the maximum of failure probability in each of the bounds in eq.~(\ref{concen2}). Without loss of generation, we set $\epsilon_{\rm c} = \epsilon_{\rm PA} =\epsilon =\epsilon_{\rm a} = 10^{-10}$.

With the concentration inequalities, we can estimate the upper bound of the phase error rate $\overline{E}_{\rm p}$. Assuming that Bob
is the internal eavesdropper, the raw key bits are divided into two parts based on Alice’s basis. The raw key bits in the X
basis are used to form secure key bits and estimate the bit error rate and the raw key bits in the Y basis are used to bound the
phase error rate. First, we offer the number of bit errors and the number of detection events in the Y basis and then record them as $m_y$ and $n_y$. In addition, the number of detection events in the X basis is also offered and recorded as $n_x$. Then, we use the concentration inequality to obtain the upper bound ${m'_y}$ on the expected number of bit errors in the Y basis, with which we find the upper bound ${E_b^Y}'={m'_y}/n_y$ on the expected bit error rate in the Y basis. Based on eq.~(\ref{eq3}), the expected phase error rate in the X basis $E'_{\rm p}$ can be bounded by ${E_b^Y}'$. Then we find the expected number of phase errors in the X basis ${m'_{\rm p}}=n_x {E'_{\rm P}}$. Next, we use the concentration inequality to calculate the upper bound $\overline{m}_{\rm p}$ on the observed number of phase errors in the X basis. Then we can derive $\overline{E}_{\rm p}$ with the formula $\overline{E}_{\rm p}=\overline{m}_{\rm p}/n_x$.


%

\end{document}